%% file: main.tex
\documentclass[journal]{IEEEtran}

\usepackage{amsmath}
\usepackage{multicol}
\usepackage{mathtools}
\usepackage{dsfont}
\usepackage{subcaption}
\usepackage{booktabs}
\usepackage{hyperref}
\usepackage[dvipsnames]{xcolor}
\usepackage{soul}
\sethlcolor{yellow}

\usepackage{xcolor}

\usepackage{wrapfig}

\usepackage{fancyhdr}
\fancypagestyle{firstpage}{%
\chead{\small This article has been accepted for publication in IEEE Transactions on Wireless Communications. This is the author's version which has not been fully edited and
content may change prior to final publication. Citation information: DOI 10.1109/TWC.2022.3189604}

\cfoot{\small This work is licensed under a Creative Commons Attribution 4.0 License. For more information, see https://creativecommons.org/licenses/by/4.0/}
}

\begin{document}

\title{Ising Machines' Dynamics and Regularization for Near-Optimal MIMO Detection}

\author{\IEEEauthorblockN{Abhishek~Kumar~Singh$^{1,2,3}$, Kyle~Jamieson$^1$, Peter~L.~McMahon$^4$ and Davide~Venturelli$^{2,3}$}\\
\IEEEauthorblockA{$^1$\textit{Department of Computer Science, Princeton University}\\$^2$\textit{USRA Research Institute for Advanced Computer Science}\\$^3$\textit{Quantum AI Laboratory (QuAIL), NASA Ames Research Center}\\$^4$\textit{School of Applied and Engineering Physics, Cornell University}}}

\maketitle
\thispagestyle{firstpage}

\begin{abstract}
Optimal MIMO detection is one of the most computationally challenging tasks in wireless systems. We show that new analog computing approaches, such as Coherent Ising Machines~(CIMs), are promising candidates for performing near-optimal MIMO detection. We propose a novel regularized Ising formulation for MIMO detection that mitigates a common error floor issue in the naive approach and evolve it into a regularized, Ising-based tree search algorithm that achieves near-optimal performance. By means of numerical simulation using the Rayleigh fading channel model, we show that in principle, a MIMO detector based on a high-speed Ising machine (such as a CIM implementation optimized for latency) would allow a higher transmitter antennas (users)-to-receiver antennas ratio and thus increase the overall throughput of the cell by a factor of two or more for massive MIMO systems. Our methods create an opportunity to operate wireless systems using more aggressive modulation and coding schemes and hence achieve high spectral efficiency: for a $16\times16$ MIMO system, we estimate around 2.5$\times$ more throughput in the mid-SNR regime ($\approx 12$~dB) and 2$\times$ more throughput in the high-SNR regime~($>$20~dB) as compared to the industry standard, a Minimum-Mean Square Error~(MMSE) linear decoder.
\end{abstract}

\IEEEpeerreviewmaketitle

\input{intro_related}
\input{isingFormulation}
\input{cim}

\input{ri-mimo}
\input{trim}

\input{evaluation}
\input{conclusion}

\input{acknowledgements}

\bibliographystyle{IEEEtran}
\bibliography{IEEEabrv,reference}
\include{biography}
\end{document}

%% file: intro_related.tex
\section{Introduction}
Wireless technologies have recently undergone tremendous growth in terms of supporting more users and providing higher spectral efficiency, with the next generation of cellular networks planning to support massive machine-to-machine communication~\cite{mmtc}, large IoT networks~\cite{nbiot}, and unprecedented data rates~\cite{5gdataRate}. The number of mobile users and data usage is rapidly increasing~\cite{IMT}, and while data traffic has been predominantly downlink, the volume of uplink traffic is becoming ever higher~\cite{uplinkTrafficExplosion} due to the emergence of interactive services and applications. The problem of optimal and efficient wireless signal detection in a Multiple-Input,~Multiple-Output (MIMO) system is central to this rapid growth and has been a key interest of network designers for several decades. While the optimal Maximum Likelihood (ML) detector is well known, it attempts to solve an NP-Hard problem~\cite{mimoReview} exactly, and so its implementation is usually impractical and infeasible for real-world systems. These computational challenges have prompted network designers to seek optimized implementations such as the Sphere Decoder~\cite{sphere,sphereIeee}, or sub-optimal approximations with polynomial complexity. These methods include linear detectors (like the Minimum Mean Square Error decoder~(MMSE)~\cite{Tse05fundamentalsof}) which perform channel inversion, successive interference cancellation (SIC) based techniques~\cite{sicMimo1} that focus on decoding each user sequentially while canceling inter-user interference, approximate tree search algorithms like the Fixed Complexity Sphere Decoder~(FSD)~\cite{fcsd} and Lattice-reduction based algorithms which involves pre-processing the channel to produce a reduced lattice basis~\cite{latRedux}. However, even today, practical methods that achieve near-optimal performance for systems with large numbers of both users and antennas are lacking~\cite{mimoSurveyLarge}. 

Instead, Massive MIMO systems~\cite{lundMM,megaMIMO}, where the number of base station antennas is much larger than the number of users, have emerged as the dominant solution to the poor bit error performance of practically-feasible MIMO detectors. Such systems have extremely well-conditioned channels, and hence even linear detectors like MMSE achieve near-optimal BER performance~\cite{mmseMassiveMimo1,mmseMassiveMimo3}. While this enables very low BER without any significant computational load, unfortunately, it comes at the cost of extreme under-utilization of available radio resources, as we can potentially support much larger spectral efficiency by concurrently serving more uplink users~(up to the number of base station antennas). The existence of a near-optimal MIMO detector with practical complexity for large MIMO systems would enable the expansion of massive MIMO systems to serve more users by having a base station antenna-user ratio of less than two, improving the overall uplink throughput of the system several-fold.

In the Computer Architecture and Physics communities, the last decade has seen a rise of a novel class of analog computers that use the dynamics of a physical system to heuristically find solutions to optimization problems framed as instances of the \emph{Ising model}, one of the most studied frameworks for magnetism in statistical mechanics. These methods include Quantum Annealing~\cite{kim2019leveraging,hauke2020perspectives}, Coherent Ising Machines (optical or opto-electronic systems)~\cite{wang2013coherent,marandi2014network,mcmahon2016fully,inagaki2016coherent,dopo,oeo}, and Oscillator-based Ising Machines~\cite{oim}. These already show promise as practical computational structures for addressing some NP-hard problems arising in practical applications. The application of Quantum Annealing~(QA) and Ising machines to MIMO detection is starting to be investigated in the last couple of years~\cite{review2021toappear,QAdeluna} and has shown promising results. The QuAMax MIMO detector~\cite{minsung} leverages quantum annealing for MIMO detection. A classical-quantum hybrid approach to QA-based ML-MIMO was proposed in~\cite{pSuccessQA}. QA has also shown promising results for other tough computational problems in wireless systems like Vector Perturbation Precoding~\cite{qavp}. In~\cite{minsungParallelTemp}, authors explore the use of Parallel Tempering for Ising-based MIMO detection (ParaMax), improving the performance of QuAMax. However, these recent works~\cite{minsung,minsungParallelTemp} that leverage a straightforward mapping of ML-MIMO decoding problem to the Ising model experiences an error floor in the bit error rate~(BER) versus the signal-to-noise ratio (SNR) characteristics.

\begin{figure*}[ht]
    \centering
    \includegraphics[width=\textwidth]{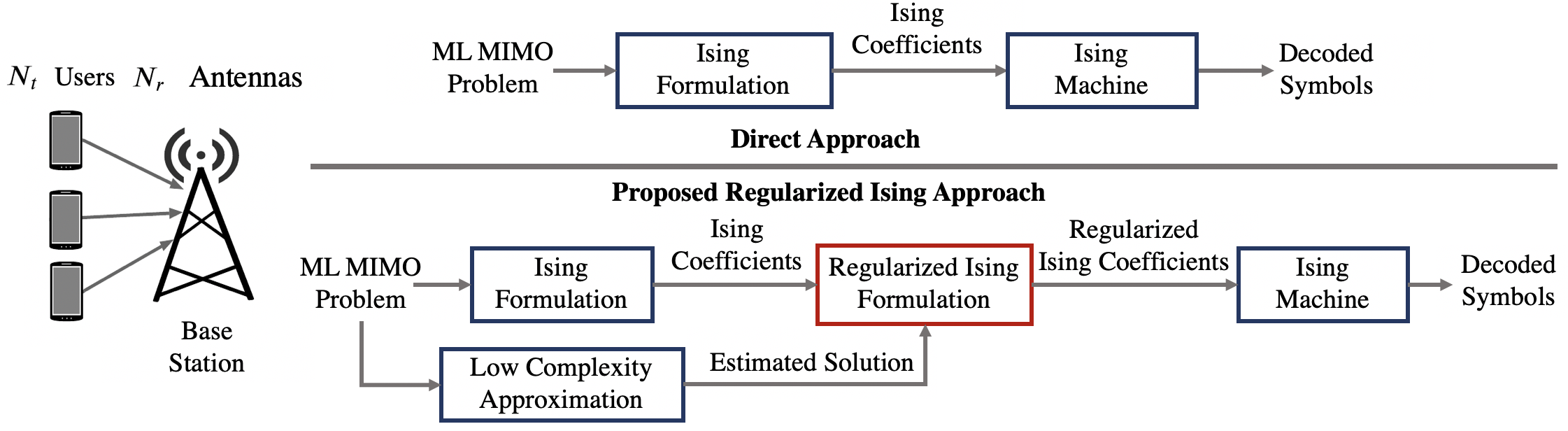}
    \caption{Uplink Maximum Likelihood MIMO detection (ML-MIMO) using Ising Machines, illustrating the differences between the proposed regularised Ising approach and the direct application of the Ising formulation.} %like QuAMax~\cite{minsung}.}
    \label{fig:riMimoIntro}
\end{figure*}

In this paper, we observe that this error floor is present in the regime of practical deployment for MIMO detection. More specifically, in the regime relevant for real systems (uncoded BER of $10^{-3} - 10^{-6}$), even if we dismiss the limitations of non-idealized physics-based Ising solvers, depending on the SNR, there are many interesting scenarios in which they would not serve as good MIMO detectors in practical systems if the known Ising formulation of the ML-MIMO detection problem is used. Hence we propose a novel regularized Ising formulation of the ML-MIMO problem, using a low-complexity approximation, that significantly improves the BER performance of the existing state-of-the-art~(see Fig.~\ref{fig:riMimoIntro} for an overview of our decoder architecture) and show that it mitigates the error-floor problem present in the existing state-of-the-art Ising machine-based MIMO detectors.

Our proposed methods improve upon the drawbacks of several popular MIMO detectors: the MMSE decoder has very low complexity but has extremely bad BER performance, and our proposed methods can vastly outperform MMSE by achieving near-optimal BER performance. While SIC-based and FSD-based methods can provide better BER than MMSE, SIC requires a sequential implementation, and FSD has limited parallelism; hence, they cannot fully utilize the large pool of parallel computing resources available in the FPGA/ASIC/GPU-assisted BS implementations. In contrast, our proposed methods have a very high degree of parallelism and can efficiently utilize the large pool of parallel computing resources to meet the processing requirements of state-of-the-art LTE systems.

It is important to note that while we discuss CIMs in this paper, there are a number of other promising Ising-machine technologies~\cite{vadlamani2020physics}, including quantum annealing~\cite{kim2019leveraging,hauke2020perspectives}, photonic Ising machines besides CIMs~\cite{roques2020heuristic,prabhu2020accelerating,babaeian2019single,pierangeli2019large}, digital-circuit Ising solvers~\cite{yamaoka201520k,aramon2019physics,goto2019combinatorial,goto2021high,leleu2020chaotic}, bifurcation machines~\cite{tatsumura2021scaling}, oscillation based Ising machines~\cite{oim}, and memristor and spintronic Ising machines~\cite{sutton2017intrinsic,grollier2020neuromorphic,cai2020power}. Note that, another promising approach of using Quantum Approximate Optimization Algorithm~(QAOA) for solving the maximum likelihood detection problem~\cite{jingjing} has been proposed as well; however, current evaluation and bench-marking is limited to Single-Input-Single-Output~(SISO) channels and very small MIMO scenarios.\\ 
The main contributions of this paper are as follows:
\begin{itemize}
    \item We propose a novel regularised Ising~(RI) formulation that enhances the probability of finding the ground state and provides significant gains in BER performance over the existing state-of-the-art. We show that it mitigates the error-floor problem present in the existing state-of-the-art Ising machine-based MIMO detectors.
    \item We propose Regularised Ising MIMO~(RI-MIMO), which is based on the proposed RI formulation, and show that it is asymptotically optimal and can achieve near-optimal performance, in the relevant BER regime (an uncoded BER of $10^{-3}$-$10^{-6}$), within practical complexities.
    \item We further evolve RI-MIMO into Tree search with RI-MIMO~(TRIM) that allows us to achieve better performance when the complexity of the underlying MIMO detection problem increases with higher-order modulations.
    \item Finally, we estimate how a physical implementation of a CIM could employ these algorithms to bridge the gap between optimality and efficiency in MIMO detection and may realize a near-optimal MIMO detector, meeting the cellular-system-complexity and timing requirements of practical scenarios, years before quantum-annealing technology would reach the necessary maturity~\cite{kasi2021challenge,sankar2021benchmark}. 
\end{itemize}
The rest of the paper is organized as follows. Section~\ref{sec:IsingMIMO} describes the MIMO system model and the reduction of the ML-MIMO problem to an Ising optimization problem. Section~\ref{sec:CIM} is a primer on CIMs. Section~\ref{sec:riMimo} describes our novel Ising formulation and the proposed MIMO detection algorithms (RI-MIMO and TRIM). Section~\ref{sec:eval} contains an extensive evaluation of BER and spectral efficiency of our methods. We show that our techniques mitigate the error floor problem and can achieve the same BER as the Sphere Decoder for $16 \times 16$ MIMO with BPSK modulation. We study the impact of practical constraints, like finite precision, on the performance of CIM for MIMO detection. We perform extensive empirical experimentation for parameter tuning. We show that for higher-order modulations, our methods provide a 10-15~dB gain over the MMSE receiver. We evaluate the spectral efficiency of our methods with Adaptive Modulation and Coding (AMC) and show that our methods can achieve around 2.5$\times$ more throughput in the mid-SNR regime ($\approx$12 dB) and 2$\times$ more throughput in high-SNR regime($>$20 dB) than the industrially-viable MMSE decoding. We evaluate the BER and spectral efficiency of our methods for several massive MIMO systems~($16\times32$, $32\times64$, $32\times128$, $64\times128$ and $64\times256$). We also evaluate our methods on the geometry-based QuaDRiGa~\cite{quadriga} channel model. Finally, we conclude in Section \ref{sec:conclusions}. 

%% file: isingFormulation.tex
\section{System Model}
\label{sec:IsingMIMO}
In this section, we will describe the MIMO system model, the MIMO Maximum Likelihood Detection (ML-MIMO) problem, and the transformation between the ML-MIMO problem and its equivalent Ising problem. 

Consider the UL transmission in a MIMO system~\cite{mimoSysModel} with $N_r$ antennas at the base station (BS) and $N_t$ users, each with a single antenna. $\mathbf{x_o} = \{x_1,x_2,...x_{N_t}\}^ T$ is the transmit vector where $x_i$ is the symbol transmitted by user $i$. $\mathbf{y} = \{y_1,y_2,...y_{N_r}\}^ T$ is the received vector where $y_j$ is the signal received by antenna $j$. Each $x_i$ is a complex number drawn from a fixed constellation $\Omega$. The channel between user $j$ and receive antenna $i$ is expressed as a complex number $h_{ij}$ that represents the channel's attenuation and phase shift of the transmitted signal $x_j$. Let $\mathbf{H}$ denote the complex-valued channel matrix,
\begin{equation}
    \mathbf{y} = \mathbf{H}\mathbf{x_o} + \mathbf{n},
    \label{eq:MIMO}
\end{equation}
where $\mathbf{n}$ denotes Additive White Gaussian Noise (AWGN). With AWGN, the optimal receiver is the Maximum Likelihood receiver~\cite{sphereIeee} which is given by
\begin{equation}
    \mathbf{\hat{x}_{ML}} = \arg \min_{\mathbf{x} \in \Omega^{N_{t}}} \|\mathbf{y} - \mathbf{H}\mathbf{x}\|^2
    \label{eq:ML-MIMO}
\end{equation}

An Ising optimization problem~\cite{minsung} is quadratic unconstrained optimization problem over $N$ spin variables, as follows:
\begin{eqnarray}
 \arg \min_{s_1,s_2,...s_N} -\sum_{i=1}^N h_{i}s_{i} - \sum_{i \neq j}J_{ij}s_{i}s_{j}\nonumber\\=\arg \min_{\mathbf{s}\in \{-1,1\}^N} -\mathbf{h}^T\mathbf{s} - \mathbf{s}^T\mathbf{J}\mathbf{s},
    \label{eq:Ising}
\end{eqnarray}
where each spin variable $s_{i} \in \{-1,1\}$, or in its vector form (RHS)  $\mathbf{s} = \{s_1,s_2,...s_N\}$, 
where all diagonal entries of the matrix  $\mathbf{J}$ are zeros.

The minimization problem expressed in (\ref{eq:ML-MIMO}) can be equivalently converted into Ising form by expressing $\mathbf{x}$ using spin variables. 
The first step is derive a real valued equivalent of (\ref{eq:MIMO}), which is obtained by the following transformation,
\begin{equation}
\mathbf{\Tilde{H}} = 
  \left[ {\begin{array}{cc}
   \Re(\mathbf{H}) & -\Im(\mathbf{H}) \\
   \Im(\mathbf{H}) & \Re(\mathbf{H}) \\
  \end{array} } \right],\text{~}
\ifdefined\isArxiv
\end{equation}
\begin{equation}
\fi
    \mathbf{\Tilde{y}} =
  \left[ {\begin{array}{c}
   \Re(\mathbf{y}) \\
   \Im(\mathbf{y}) \\
  \end{array} } \right],\text{~}\mathbf{\Tilde{x}}=
  \left[ {\begin{array}{c}
   \Re(\mathbf{x}) \\
   \Im(\mathbf{x}) \\
  \end{array} } \right],
  \label{realTransVec}
\end{equation}

The ML receiver described in (\ref{eq:ML-MIMO}) has the same expression under the transformation and the optimization variable $\mathbf{\tilde{x}}$ is real valued. Let us say $\mathbf{\tilde{x}}$ has $N$ elements, which are drawn from a square M-QAM constellation, then each element of the optimization variable $\mathbf{\tilde{x}}$ takes integral values in the range $\Omega_r= \{-\sqrt{M}+1,-\sqrt{M}+3,...\sqrt{M}-1 \}$. The number of bits needed to express $\Omega_r$ are given by $r_b = \lceil \log_{2}(\sqrt{M})\rceil$. let $\mathbf{s}$ be an $N*r_b\times1$ spin vector such than each element of $\mathbf{s}$ can take values $\{-1,1\}$. Then, element $j$ of $\mathbf{\tilde{x}}$ can be represented using $r_b$ spin variables $\{s_j,s_{j+N}...s_{j+(r_b-1)N}\}$,
\begin{equation}
\label{eq:elementBitRep}
    \tilde{x}_j = \sum_{i=1}^{r_b}2^{r_b-i}(s_{j+(i-1)N}+1) - (\sqrt{M} - 1)
\end{equation}
Let us define the transform matrix $\mathbf{T} = \left[2^{r_b-1}\mathds{I}_{N}\text{~~~}2^{r_b-2}\mathds{I}_{N}\text{~}...\text{~}\mathds{I}_{N}\right]$,
then $\mathbf{\tilde{x}}$ can be expressed as,
\begin{equation}
    \mathbf{\tilde{x}} = \mathbf{T}(\mathbf{s}+\bar{\mathds{1}}_{N*r_b}) - (\sqrt{M}-1)\bar{\mathds{1}}_{N}
    \label{eq:qamSpinTrans}
\end{equation}
For a rectangular QAM constellations and BPSK, the $\Re(\mathbf{x})$ and $\Im(\mathbf{x})$ in (\ref{realTransVec}) have different range and (\ref{eq:elementBitRep}) can be accordingly modified to construct the transform matrix $T$. We substitute (\ref{eq:qamSpinTrans}) in the real valued ML problem to obtain the ising formulation for ML receiver. Let $\mathbf{z} = \mathbf{\tilde{y}} - \mathbf{\tilde{H}}\mathbf{T}\bar{\mathds{1}}_{N*r_b} + (\sqrt{M}-1)\mathbf{\tilde{H}}\bar{\mathds{1}}_{N}$, then the Ising problem for ML-MIMO is described by,
\begin{equation}
    \mathbf{h} = 2*\mathbf{z}^{T}\mathbf{\tilde{H}}\mathbf{T}\text{,~~~~}\mathbf{J} = -\textit{zeroDiag}(\mathbf{T}^{T}\mathbf{\tilde{H}}^T\mathbf{\tilde{H}T}),
\end{equation}
where $\textit{zeroDiag}(\mathbf{W})$ sets the diagonal elements of matrix $W$ to zero. We further scale the problem such that all the coefficients lie in $[-1,1]$. The Ising solution can be converted to real valued ML solution using  (\ref{eq:qamSpinTrans}), which can be then converted to the complex valued solution for the original problem described in (\ref{eq:ML-MIMO}) by inverting the transform described by (\ref{realTransVec}). 

%% file: cim.tex
\section{Coherent Ising Machines (CIM)}
\label{sec:CIM}
In simple terms, an Ising Machine can be described as a module that takes an Ising Problem (Eq.~\ref{eq:Ising}) as input and outputs a candidate solution, according to an unknown probability distribution that depends on a few parameters. We refer to a single, independent run on the Ising Machine as an "anneal", borrowing nomenclature from the simulated/quantum-annealing methods. After each run, the machine is reset. 
A common approach is to run several samples of a single Ising problem instance and then return the best-found solution (the "ground state") in the sample. 

\emph{Coherent Ising Machines} (CIMs), as originally conceived~\cite{marandi2014network,mcmahon2016fully,inagaki2016coherent}, implement the search for the ground state of an Ising problem by using an optical artificial spin network. Their baseline architecture encodes the Ising spins into a train of time-resolved, phase-coherent laser pulses traveling on an optical fiber loop, undergoing controlled interference between all pairs of wavepackets.  
The phase dynamics of the pulses is governed by the presence of a non-linear element in the form of a degenerate-optical-parametric-oscillator (DOPO). For the purpose of our work, this version of the CIM can be modeled using a system of stochastic differential equations~\cite{wang2013coherent}. In this work, we will use a model of the CIM that describes its continuous-time limit evolution neglecting quantum effects, which has been proven to be fitting the experiments of multiple devices and represents a baseline setup for more sophisticated embodiments. Following Ref.~\cite{dopo}, the in-phase ($c_{i}$) and quadrature ($q_{i}$) components of each signal-variable can be modeled using the following differential equations:

\begin{eqnarray}
       dc_{i} &=& [(-1 + p - c_{i}^2 - q_{i}^2)c_{i} + 
       C\sum_{j}J_{ij}c_{j}]dt\nonumber\\
       &+& \frac{1}{A_{s}}\sqrt{c_{i}^2 + q_{i}^2 + \frac{1}{2}}dW_{1}\\
    dq_{i} &=& (-1 - p - c_{i}^2 - q_{i}^2)q_{i}dt + \frac{1}{A_{s}}\sqrt{c_{i}^2 + q_{i}^2 + \frac{1}{2}}dW_{2}\nonumber
\label{eq:dopo}
\end{eqnarray}
\noindent where the normalized pump rate ($p$) are CIM parameters that relates to the laser used in the machine and can be tuned easily. The constant C is typically fixed by design considerations (mostly by the power transmission coefficient and the laser saturation amplitude). $J_{ij}$ is the Ising coupling coefficient from the $j^{th}$ pulse to the $i^{th}$ pulse, which is programmable. The stochasticity is introduced through $dW_1$ and $dW_2$, which are independent Gaussian-noise processes. The variable $t$ is time (normalized with respect to the photon decay rate). An Ising problem with the spin-spin-coupling matrix $J$ is encoded in the CIM by setting the optical couplings $\Tilde{\zeta}_{ij} \propto J_{ij}$. The anneal consists of pumping energy into the system by gradually varying $p$. Heuristically this is implemented in a schedule at a speed that is some monotonic function of $N$. The solution to the  Ising problem is read out at the end of the anneal by measuring the in-phase component of each DOPO $c_{i}$, and interpreting the sign of each as a spin value $s_{i}$, i.e. $s_i = \textrm{sign}(c_i)$.
\subsection{CIM Simulator}
\label{sec:CIMsim}
To enable the study of how an ideal CIM would perform on solving Ising instances related to the application at hand (MIMO detection), we implemented a software simulator of a CIM that integrates the differential equations described in ($\ref{eq:dopo}$), using double precision numerics implemented in MATLAB. The annealing schedule consists in varying the pump parameter as $p(t)=2*\tanh(\frac{2t}{N})$. For the purpose of numerical integration, $dt = 0.01$ and the total anneal time is $128 * dt$, which corresponds to 128 steps of numerical integration). In both numerics and experiments, we can also sample the state of the CIM at intermediate times (number of measurements $N_m$) to retrieve more candidate solutions for the programmed Ising problem and then select the best solution found; so each run actually evaluates $N_m=128$ candidate solutions. However, most of the time, the best solution corresponds to the one at the final measurement. Following the inspiration from design described in Ref.~\cite{dopo} we set $C=\sqrt{10}$. As the pump rate is gradually increased from 0 to 2, the amplitudes of DOPO pulses demonstrate a bifurcation and settle to either a positive (indicating spin +1) or a negative value (indicating spin -1).

%% file: ri-mimo.tex
\section{Design}
\label{sec:riMimo}
In this section, we propose the RI-MIMO detector, based on our novel regularised Ising formulation of maximum-likelihood MIMO receiver, which mitigates the error floor problem. We use a single auxiliary spin variable to transform the Ising problem into a form compatible with CIMs. We finally propose TRIM, a novel tree search algorithm based on RI-MIMO, which enhances the performance for higher-order modulations. 

\subsection{RI-MIMO: Regularized Ising-MIMO}

 As noted before, when we use Ising machines for MIMO detection, there is a finite probability of not returning the ground state of the problem, even at zero noise, and hence, we cannot decrease the BER beyond a certain limit (i.e., there is an "error floor"). We can observe this error floor in existing works on ML-MIMO detection using a Quantum Annealer \cite{minsung}. In this section, we present Regularized Ising-MIMO~(RI-MIMO) that mitigates the error floor and provides significant performance enhancements. 
\subsubsection{Regularization}
\textit{The key idea is to add a regularisation term based on a low complexity estimate of the solution, which, as we will see later, will enhance the probability of finding the ground state of the Ising problem and hence improve the BER performance}. The maximum likelihood MIMO receiver is given by (\ref{eq:ML-MIMO}). Let us say that we have a polynomial-time estimate (obtained by algorithms like MMSE or ZF) $\mathbf{x}_P$.
Let $\mathbf{s}_P$ be the spin vector corresponding to $\mathbf{x}_P$ obtained from (\ref{eq:qamSpinTrans}). We add to the Ising form a penalty term for deviations from the poly-time estimate, which would penalize non-optimal solutions in low noise scenarios, to obtain:
\begin{eqnarray}
      \mathbf{\hat{s}} &=& \arg\min_{\mathbf{s}\in \{-1,1\}^N} -\mathbf{h}^T\mathbf{s} - \mathbf{s}^T\mathbf{J}\mathbf{s} + r(\rho,M,N_t)\|\mathbf{s} - \mathbf{s}_P\|^2\nonumber\\
       &=& \arg\min_{\mathbf{s}\in \{-1,1\}^N} -(\mathbf{h} +2r(\rho,M,N_t)\mathbf{s}_P)^T\mathbf{s} - \mathbf{s}^T\mathbf{J}\mathbf{s}, \label{eq:ri-mimo}
\end{eqnarray}
where $r(\rho,M,N_t)$ is a regularization parameter dependent of the SNR, modulation and number of users. This style of regularisation falls in the class of generalized Tikhonov regularisation~\cite{tikhonov}. We will look at the choice of $r(\rho,M,N_t)$ in Section~\ref{sec:regFactor}.

The \textit{RI-MIMO-$N_a$} algorithm is then defined as follows:
\begin{enumerate}
    \item Convert the ML-MIMO detection problem into the Ising form as described in Section~\ref{sec:IsingMIMO}. 
    \item Add the regularisation term as described by (\ref{eq:ri-mimo}).
    \item Perform $N_a$ anneals using an Ising machine.
    \item Select the best solution from the candidate solutions ($\mathbf{x_1},\mathbf{x_2}...\mathbf{x}_{N_a\times N_m}$) generated by the Ising machine (one from each measurement times the number of anneals $N_m\times N_a$) and the MMSE solution~($\mathbf{x_{mmse}}$). The cost associated with any given solution $\mathbf{w}$ is given by ML-MIMO cost function ($\|\mathbf{y} - \mathbf{H}\mathbf{w}\|^2$) and the best solution is given by 
    \begin{equation*}
        \arg \min_{\mathbf{w} \in \{\mathbf{x_1},\mathbf{x_2},...\mathbf{x_{N_a \times N_m},\mathbf{x_{mmse}}\}}} \|\mathbf{y} - \mathbf{H}\mathbf{w}\|^2
    \end{equation*}
    
\end{enumerate}

In contrast to RI-MIMO, a trivial way to lower the error floor is to increase the number of runs per instance of the problem. However, each additional anneal increases the overall computation time as well. Another simple approach is to use a direct combination of Ising machine-based ML-MIMO and MMSE, where we select the best solution among those generated by the Ising machine and the MMSE solution. This removes the error floor but, at higher SNR, the reduction in BER is primarily due to the MMSE receiver and, as we will show later, our proposed techniques have much better BER performance.
\subsection{Solution of an Ising problem having access only to programmable quadratic couplings}
\label{sec:biasCIM}
Not all CIMs are designed to solve Ising problems containing a bias term ($\mathbf{h}^T\mathbf{s}$ in (\ref{eq:Ising})). In order to solve a general Ising problem using a CIM that does not natively support bias terms (although some, such as the implementation used in \cite{mcmahon2016fully}, do), we introduce an auxiliary spin variable $s_a$ and solve the following  Ising problem:
\begin{equation}
    \min_{s_a,s_1,s_2,...s_N} -\sum_{i=1}^N h_{i}s_{i}s_a - \sum_{i \neq j}J_{ij}s_{i}s_{j},
    \label{eq:auxIsing}
\end{equation}
which contains no bias terms and can be solved using an Ising machine that doesn't support bias terms. (\ref{eq:auxIsing}) has two degenerate solutions: $[\{\hat{s_{i}}\}_{i=1}^N, \hat{s}_{a} = 1]$ and  $[\{-\hat{s_{i}}\}_{i=1}^N, \hat{s}_{a} = -1]$. Note that $\{\hat{s_{i}}\}_{i=1}^N$ is the solution for the original Ising problem (\ref{eq:Ising}). Hence we can obtain the solution to the original Ising problem from the solutions of the auxiliary Ising problem.

\subsubsection{Proof}
\label{sec:proofIsing}
The Ising problem we intend to solve is given by,
\begin{equation}
     [\mathbf{\hat{s}},J(\mathbf{\hat{s}})] = \min_{\mathbf{s}\in \{-1,1\}^N} -\mathbf{h}^T\mathbf{s} - \mathbf{s}^T\mathbf{J}\mathbf{s},
 \end{equation}
 where $\mathbf{\hat{s}}$ is an optimal solution. The auxilary Ising problem is given by,
 \begin{equation}
      [(\mathbf{\bar{s}},\hat{s}_a),J_a(\mathbf{\bar{s}},\hat{s}_a)] = \min_{\mathbf{s}\in \{-1,1\}^N\text{,~}s_a\in\{-1,1\}} -(\mathbf{h}^T\mathbf{s})s_a - \mathbf{s}^T\mathbf{J}\mathbf{s},
 \end{equation}
 where $(\mathbf{\bar{s}},\hat{s}_a)$ denote an optimal solution for the auxiliary Ising problem. Note both $\mathbf{\bar{s}}$ and $\mathbf{\hat{s}}$ are $N\times1$ spin vectors. If $\hat{s_{a}} = 1$ then, $J_a(\mathbf{\bar{s}},1)$ = $J(\mathbf{\hat{s}})$. Let us assume to the contrary:\\
 1. $J_a(\mathbf{\bar{s}},1) < J(\mathbf{\hat{s}})$, then $J(\mathbf{\bar{s}}) < J(\mathbf{\hat{s}})$ which contradicts $\mathbf{\hat{s}}$ is optimal.\\ %solution for the original ising problem.\\
 2. $J_a(\mathbf{\bar{s}},1) > J(\mathbf{\hat{s}})$, the $J_a(\mathbf{\bar{s}},1) > J_a(\mathbf{\hat{s}},1)$ which contradicts $(\mathbf{\bar{s}},1)$ is optimal.% solution for auxiliary ising problem.

Hence, $J_a(\mathbf{\bar{s}},1)$ = $J(\mathbf{\hat{s}})$. Note that, the auxiliary Ising problem has two degenerate solutions $(\mathbf{\bar{s}},1)$ and $(-\mathbf{\bar{s}},-1)$, where $\mathbf{\bar{s}}$ is also an optimal solution for the original Ising problem. Therefore, we can obtain the solution to the original problem from the solutions of the auxiliary problem.

%% file: trim.tex
\subsection{TRIM: \textbf{T}ree search with \textbf{R}egularized \textbf{I}sing-\textbf{M}IMO}
\ifdefined\isArxiv
\begin{figure*}[t]
\centering
\includegraphics[width=\linewidth]{figure/RI-MIMO-BER-2-Arxiv.PNG}
 \caption{Bit Error Rate (BER) Curves for (Left) 16$\times$16, (Center) 20$\times$20, (Right) 24$\times$24 MIMO and BPSK modulation, illustrating the error floor problem and performance of all the tested solvers. The curves are computed over $\approx$25K MIMO instances (128 channel instances, 198 transmit vectors per channel).}
\label{fig:16x16_RI_OIM_QA}
\end{figure*}
\fi
A tree search algorithm for ML-MIMO receiver represents the process of finding the optimal solution by visiting various leaf nodes of a suitably constructed search tree. For an M-QAM constellation, there are M possible symbols for each user. We represent the M possibilities for user-1 by M nodes at depth 1.
Each of these nodes has M children, which represent the M possibilities for user 2, and so on. Any path from the root to a leaf node will represent a candidate solution for the ML-MIMO problem. As noted before, there exist several techniques that involve traversing this search tree to obtain good quality solutions in polynomial complexity.

In this section, we build on the ideas of RI-MIMO, to further improve the performance of CIM based ML-MIMO receiver by proposing a hybrid tree search algorithm. The maximum-likelihood MIMO receiver described in (\ref{eq:ML-MIMO}) can be expressed using a QR decomposition, $\mathbf{H} = \mathbf{QR}$,
\begin{equation}
    \mathbf{\hat{x}_{ML}} = \arg\min_{\mathbf{x} \in \Omega^{N_{t}}} \|\mathbf{w} - \mathbf{R}\mathbf{x}\|^2,
    \label{eq:QR-ML-MIMO}
\end{equation}
where $\mathbf{w} = \mathbf{Q}^{\dagger}\mathbf{y}$. If we fix the symbol corresponding to user $N_t$ to $u$, then $\mathbf{x} = [\mathbf{v}\text{~~}u]^T$, where $\mathbf{v} \in \Omega^{N_t - 1}$. Let 
%\begin{equation}
$\mathbf{R} =  
  \left[ \bar{\mathbf{R}}\text{   }\bar{\mathbf{c}} \right]$,
%  \label{realTransMat}
%\end{equation}
where $\mathbf{\bar{R}}$ is an $(N_t - 1) \times (N_t - 1)$ upper triangular matrix, $\mathbf{\bar{c}}$ is $(N_t - 1)\times 1$ and $r$ is a scalar. Substituting this is (\ref{eq:QR-ML-MIMO}) and simplifying, we obtain: 
\begin{equation}
    \mathbf{\hat{v}}_{ML} = \arg\min_{\mathbf{v} \in \Omega^{N_t - 1}} \|(\mathbf{\bar{w}}-\mathbf{\bar{c}}u) - \mathbf{\bar{R}}\mathbf{v}\|^2.
\end{equation}
This has the same form as (\ref{eq:ML-MIMO}) and can be solved using RI-MIMO techniques. The same procedure can be extended if we fix symbols corresponding to more than one user.

We use this formulation for building a tree search algorithm, \textit{TRIM}$_d$-$N_a$, augmented by \textit{RI-MIMO}-$N_a$. While solving the ML-MIMO problem, we consider all possibilities for $d$ users, which is the same as considering all nodes in the search tree up to depth ($d$). For each branch of the search tree, once we fix the symbols corresponding to $d$ users, we formulate the maximum likelihood problem for the remaining users as illustrated above, and solve it using \textit{RI-MIMO}-$N_a$, as illustrated in Section~\ref{sec:riMimo}. In the end, we select the best solution out of all the candidate solutions obtained by \textit{RI-MIMO}-$N_a$ sub-problems corresponding to various branches of the search tree. Note that for an $M$-QAM constellation, \textit{TRIM}$_d$-$N_a$ requires $M^{d}\times N_a$ anneals.

%% file: evaluation.tex
\section{Evaluation}
\label{sec:eval}
In this section, we will perform an extensive evaluation of RI-MIMO and TRIM in various scenarios. We will study the performance of RI-MIMO considering some practical constraints of Ising Machines (IM), like the finite precision available to program the Ising coefficients. We will also evaluate RI-MIMO and TRIM for various modulation schemes, MIMO sizes, and under adaptive MCS (Modulation and Coding Scheme). 

We will simulate (using MATLAB, see \ref{sec:CIMsim}) uplink wireless MIMO transmission between $N_t$ users with one transmit antenna each and a base station with $N_r$ receive antenna. We assume Rayleigh fading channel between them and additive white Gaussian noise (AWGN) at the receiver. We further assume, for simplicity, that the channel is known at the receiver and all users use the same modulation scheme. The BER of an $N_t\times N_r$ MIMO system is calculated as the mean BER of the $N_t$ independent data streams transmitted by $N_t$ users.  

\subsection{BER Performance}
\begin{figure*}
\centering
\includegraphics[width=\linewidth]{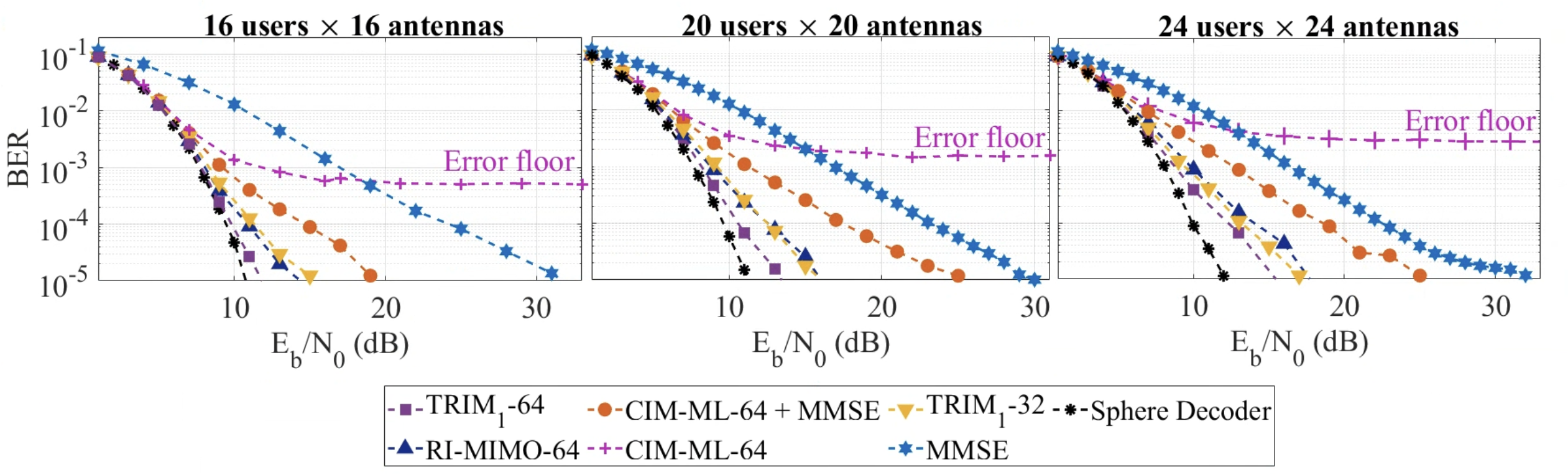}
 \caption{Bit Error Rate (BER) Curves for (Left) 16$\times$16, (Center) 20$\times$20, (Right) 24$\times$24 MIMO and BPSK modulation, illustrating the error floor problem and performance of various solvers. 
 }
\label{fig:16x16_RI_OIM_QA}
\end{figure*}

We start comparing the optimal decoder (the Sphere Decoder) and the linear MMSE decoder against \textit{RI-MIMO} and the unregularized \textit{ML-MIMO} using as a test case BPSK 16$\times$16. This case will represent a baseline for our benchmarks and their sophistication. Note that a trivial way to remove the error floor is to take the better solution out of those generated by MMSE and \textit{CIM-ML-}$M_a$. We see from Fig~\ref{fig:16x16_RI_OIM_QA} that \textit{RI-MIMO} provides much better BER than \textit{CIM-ML}, mitigating the error floor problem associated with it. We note that if we run concurrently MMSE and \textit{ML-MIMO} for each instance and we select the best of both results (\textit{CIM-ML+MMSE}), we are still less performant than \textit{RI-MIMO}. \textit{TRIM} performs similar to \textit{RI-MIMO} when both algorithms execute the same number of total anneals. (\textit{TRIM}$_1$-32 vs \textit{RI-MIMO}-64). However, as we will see later, this is not the case when higher modulations are used. The best performing algorithm is \textit{TRIM}$_1$-64, which achieves near-optimal performance for $16 \times 16$ and $20 \times 20$ MIMO with BPSK modulation. For $24\times24$ MIMO, we see that the performance gap between \textit{TRIM}$_1$-64 and Sphere Decoder increases, and a higher number of anneals are required to bridge the gap. Note that, for MIMO systems with a larger number of antennas, along with an increase in the number of anneals, increasing the depth of the tree search might be required to maintain the performance gains of TRIM.
\subsection{Complexity of RI-MIMO: From Simulation to CIM Hardware Implementation}
In this section, we explore the computational complexity of RI-MIMO and discuss the hardware considerations, for a real-world deployment of our algorithms, to satisfy the LTE requirements. While this study is performed with a simulator, our objective is to predict the performance and understand the trade-offs for designing a future hardware implementation of a CIM for ML-MIMO. To estimate the feasibility of MIMO decoding, we need to compare the computational requirements versus the perspective capabilities on a real-world perspective machine.

\textit{Infeasible scaling of FPGA/ASIC:} Before we go forward and discuss the hardware requirements of a real-world CIM, let us look at why FPGA/ASIC-based solutions will not be enough to meet the requirements of future wireless systems. As stated before, the Sphere Decoder is known to be theoretically optimal, and there are several FPGA/ASIC implementations of the Sphere Decoder in the literature; however, it is known to have exponential average complexity, (unless we operate sufficiently under the capacity or have high SNR)~\cite{sphereComp,sphereComp2}, and becomes impractical as the size of the MIMO system increases. The ETH SD~\cite{ethsd} is known to be one of the most optimized implementations of the Sphere Decoder algorithm. Using the data in \cite{ethsd}, we can approximately extrapolate the results of ETH SD by linearly extrapolating the logarithm of the average number of nodes visited (exponential scaling). As per the LTE specifications, a typical LTE deployment with 10~MHz bandwidth and 15KHz sub-carrier spacing produces 8400 MIMO detection problems (corresponding to 14 OFDM symbols per sub-frame and 600 sub-carriers) every sub-frame of 1-millisecond duration. Based on this, we compute the number of parallel ASIC/FPGA implementations of ETH SD to meet the LTE requirements (in Table~\ref{table:ethsd}), illustrating that current silicon-based implementation can only support MIMO sizes like $8\times8$ with less than 25 parallel implementations, but larger systems like $64\times64$ will require a billion parallel instances of ETH SD, which is impractical. As a result, practical deployments of large MIMO systems end up using computationally efficient linear detectors like Zero Forcing or MMSE, but their BER performance is significantly worse. The inefficient scaling FPGA implementations of Sphere Decoder and unacceptable BER performance of linear detectors prompt us to explore novel hardware/algorithmic solutions and motivate us to study CIM-based MIMO detection.
\begin{table}[h!]
\centering
\begin{tabular}{c c c} 
\toprule
\textbf{MIMO System} & \textbf{ETH SD Processing} & \textbf{Parallel ASIC requirement}\\
                     & \textbf{Throughput (Mbps)} & \textbf{for LTE (at 8 dB SNR)}\\
\hline 
4x4, 16 QAM & 20 & 7\\ 
%6x6, 16 QAM & 16 & 13\\ 
8x8, 16 QAM & 11.17 & 24\\
16x16, 16 QAM & 1.8 & 300\\
32x32, 16 QAM & 0.022& $\approx49000$\\
64x64, 16 QAM & $16.7\times10^{-7}$& $\approx$1.3 billion\\ 
\bottomrule
\end{tabular}
\caption{Parallel ASIC requirement to meet LTE processing requirements with ETH SD.}
\label{table:ethsd}
\end{table}
\begin{figure*}
    \centering
    \includegraphics[width=\linewidth]{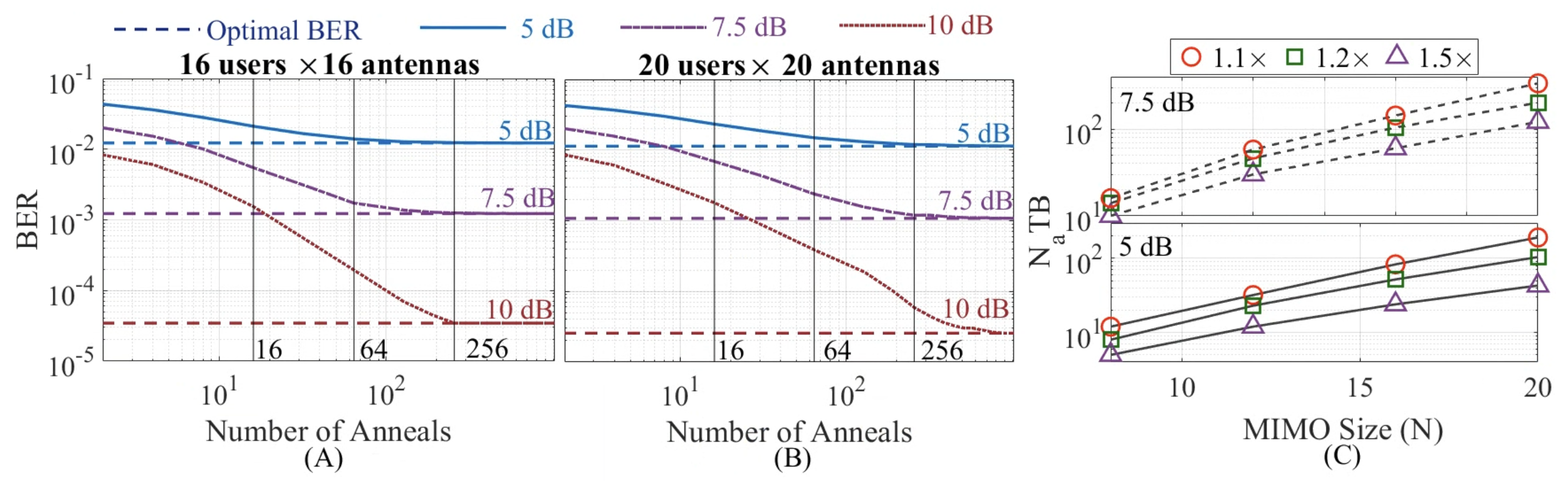}
    \caption{BER vs. $N_a$ for (A) $16\times16$, (B) $20\times20$ MIMO and BPSK modulation at 5, 7.5 and 10 dB SNR: Illustrating that the BER of RI-MIMO reduces rapidly with increase in number of anneals per instance and asymptotically approaches the optimal BER. (C) N$_a$TB vs. MIMO Size ($N$) for $N\times N$ large MIMO systems: Illustrating the exponential growth in the number of anneals required to achieve a BER which is $1.1\times$, $1.2\times$ and $1.5\times$ the optimal BER. 
    }
     \label{fig:ttb}
\end{figure*}

\begin{table*}[h!]
\centering
\begin{tabular}{l l l l l l} 
\toprule
\textbf{MIMO Order} & \textbf{Modulation} & $\mathbf{N_a}$ &  $\mathbf{N_s}$& \textbf{Throughput Requirement}\\
\hline 
Standard LTE $2\times2$ & 16 QAM & 16 &8& $1.344 \times 10^5 $ anneals per ms \\
Massive MIMO $16\times32$ & BPSK &  16 &16& $1.344 \times 10^5 $ anneals per ms \\
Large MIMO $16\times16$ & BPSK & 64 &16& $5.376 \times 10^5 $ anneals per ms \\
Large MIMO $16\times16$ & BPSK & 128 &16& $1.075 \times 10^6 $ anneals per ms \\
Large MIMO $20\times20$ & BPSK & 64  & 20 & $5.376 \times 10^5$ anneals per ms \\
Large MIMO $16\times16$ & 4 QAM &  128 &32& $1.075 \times 10^6 $ anneals per ms \\
Large MIMO $8\times8$ & 16 QAM &  256 &32& $2.15 \times 10^6 $ anneals per ms \\
Large MIMO $16\times16$ & 16 QAM &  256 &64& $2.15 \times 10^6 $ anneals per ms \\
\bottomrule
\end{tabular}
\caption{Requirements for an Ising machine (or set of Ising machines) to meet the Ising mapping and LTE processing constraints for various MIMO systems. $N_\textrm{a}$ is the number of anneals required per Ising instance to obtain a satisfactorily accurate solution. $N_\textrm{s}$ is the number of spins in each Ising instance. In this table we assume that for all MIMO orders listed, 8400 Ising instances need to be solved every 1 millisecond. \textit{Throughput Requirement} gives the number of anneals (of instances each with $N_\textrm{s}$ spins) that the set of Ising machines needs to process in a millisecond, given by $8400 \times N_\textrm{a} / 1~\textrm{ms}$.}
\label{table:freq}
\end{table*}

Returning to the discussion on hardware considerations for real-world deployment of our methods, it is important to note that the 8400 MIMO detection problems~(generated every millisecond) are independent, so different instances can be solved in an embarrassingly parallel fashion if there are enough spin variables available in the machine. In Fig.~\ref{fig:ttb}(A) and~\ref{fig:ttb}(B) , we look at the variation of BER with the Number of Anneals per instance ($N_a$) and try to characterize the value of $N_a$ required to provide near-optimal performance. We simulate the RI-MIMO-$N_a$ system, with BPSK modulation, for various values of $N_a$. In Fig.~\ref{fig:ttb}(A) and~\ref{fig:ttb}(B), we illustrate the variation of RI-MIMO BER with $N_a$ for $16\times16$ and $20\times20$ MIMO with BPSK modulation at 5, 7.5, and 10 dB. We see that at first, the RI-MIMO-BER reduces with increasing $N_a$ and asymptotically approaches the optimal BER (Sphere Decoder). Based on the evaluation of the BER performance of our proposed algorithms, as discussed in the previous section, we see that we need to study the performance of our proposed methods using a number of anneals of the order of $N_a=16-256$ depending on the modulation and size of the MIMO problem. To investigate the variation in the required number of anneals with an increase in the size of the MIMO system, we define the metric, Number-of-Anneals-to-BER (N$_a$TB), which is equal to the number of anneals required to achieve a certain multiple of the optimal BER. In Fig.~\ref{fig:ttb}(C), we plot the variation in the number of anneals required to achieve $1.1\times$, $1.2\times$, and $1.5\times$ the optimal BER as a function of the MIMO Size ($N$). We see that, there is an exponential increase in N$_a$TB with increasing MIMO size, which is consistent with the fact that the underlying decoding problem is NP-Hard. 
\begin{figure*}[h!]
    \centering
    \includegraphics[width=\linewidth]{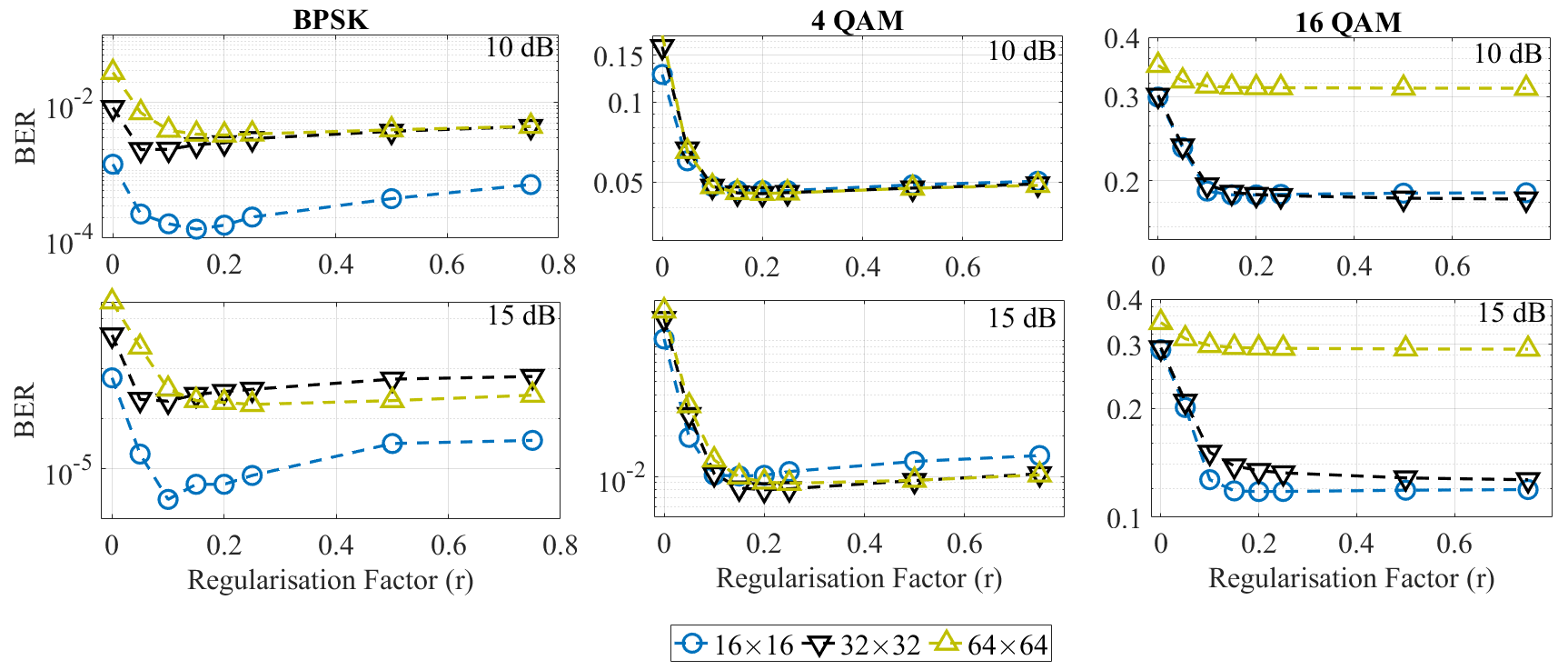}
    \caption{Bit Error Rate at 10 dB and 15 dB SNR , illustrating the performance of RI-MIMO on Coherent based Ising Machines (CIM) for various value of regularisation factor with different MIMO sizes and modulation (using 64 anneals per instance).}
    \label{fig:reg_v_numMod}
\end{figure*}
In Table~\ref{table:freq} we list the regimes of MIMO operation that are compatible with plausibly near-term Ising machines. For each regime, the table gives a throughput requirement that a set of Ising machines would need to satisfy. The fundamental requirement we assume in all the regimes is that 8400 Ising instances must be solved within 1 millisecond. Given our findings from simulations of how many anneals ($N_\textrm{a}$) need to be performed to achieve suitable MIMO performance for each regime, we can characterize the requirement for the set of Ising machines in terms of how many anneals they must perform in total within 1 ms. Note that different MIMO regimes require Ising instances with different numbers of spins ($N_\textrm{s}$) to be solved---so, for example, building a set of Ising machines capable of meeting the requirements for Large MIMO $8 \times 8$ with 16-QAM modulation is different than for Large MIMO $16 \times 16$ with 16-QAM modulation because while they both require throughput of $2.15 \times 10^6$ anneals per ms, the former requires anneals of 32-spin Ising instances and the latter requires anneals of 64-spin Ising instances. The length of a single anneal is defined in Section~\ref{sec:CIMsim} and corresponds (in discretized dynamics) to 128 updates of the spin configuration during the anneal. The results in Table~\ref{table:freq} and our discussion so far have not been specific to any particular CIM implementation and will generalize to other forms of Ising machines that have similar working mechanisms \cite{vadlamani2020physics}, including electronic oscillator-based Ising machines \cite{oim}. However, we will now briefly give some interpretation of our results that is specific to CIMs, and in particular, time-multiplexed CIMs. In time-multiplexed CIMs such as those reported in Refs.~\cite{marandi2014network,mcmahon2016fully,inagaki2016coherent,100kCIM}, the length of an anneal (128) corresponds to the number of roundtrips of pulses around an optical cavity. In Ref.~\cite{100kCIM}, pulses representing spins are separated in time in a cavity by 200 picoseconds; in a CIM system with such a pulse spacing the time required for a single roundtrip is $N_\textrm{s} \times 200~\textrm{ps}$, and hence the time required for a single anneal is $128 \times N_\textrm{s} \times 200~\textrm{ps}$. If we consider the MIMO regimes for which Ising instances with $N_\textrm{s}=64$ spins need to be solved, a single anneal will take $1.638~\mu\textrm{s}$. Therefore the number of 64-spin anneals that a single time-multiplexed CIM with 200-ps pulse spacing can perform per millisecond is $\approx 610$. Table~\ref{table:freq} shows, for example, that the MIMO regime of Large MIMO $16 \times 16$ and $16$ QAM requires a throughput of $2.15 \times 10^6$ anneals per ms. This implies that $(2.15 \times 10^6 / 610.4) \approx 3522$ copies of this type of CIM, operating in parallel, would be required to meet the requirements for this MIMO regime. Even if the roundtrip time was decreased by a factor of 20 (which is currently under experimental investigation by groups working on time-multiplexed CIMs), leading to a $20\times$ increase in the speed of a single CIM and hence a $20\times$ reduction in the number of CIMs needed to meet the throughput requirement, one would still need $\approx 176$ copies of the CIM. While it is certainly conceivable that one could construct hundreds or thousands of copies of such a 64-spin time-multiplexed CIM, especially when considering on-chip implementations using integrated photonics \cite{luke2020wafer}, it is clear that meeting the throughput requirement is a large practical engineering challenge.

Returning to a discussion applicable to Ising machines generically, it seems likely that any Ising machine---be it a CIM or any other type---will struggle to meet the throughput requirement using just a single machine solving Ising instances serially because to solve 8400 instances in 1 ms serially, one needs to solve (up to the accuracy achieved using $N_a$ anneals in a CIM) each instance in 119 ns, and even for Ising instances with only 64 spins, this seems beyond any current machine or approach. Therefore it seems likely that multiple copies of an Ising machine running in parallel will be needed, and the question turns to how many parallel Ising machines are needed and how practical is operating such a set of machines at each LTE base station? This is a major open question and challenge for the Ising-machine community at large.
 \subsection{Optimal regularisation factor for RI-MIMO and Higher Order Modulations}
\label{sec:regFactor}
\begin{figure*}[h]
    \centering
    \includegraphics[width=\linewidth]{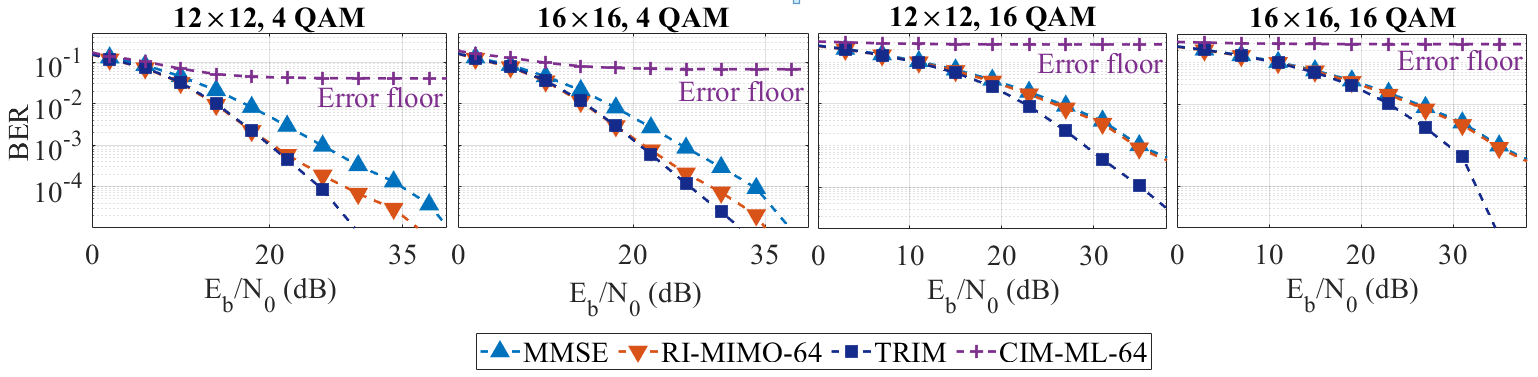}
    \caption{Bit Error Rate Curves higher order modulation schemes, illustrating the performance of RI-MIMO and TRIM on Coherent based Ising Machines (CIM). TRIM executes total 64 anneals for each instance (same as RI-MIMO-64).
    }
    \label{fig:16x16_higherMod}
\end{figure*}
\begin{figure*}
\centering
\includegraphics[width=\linewidth]{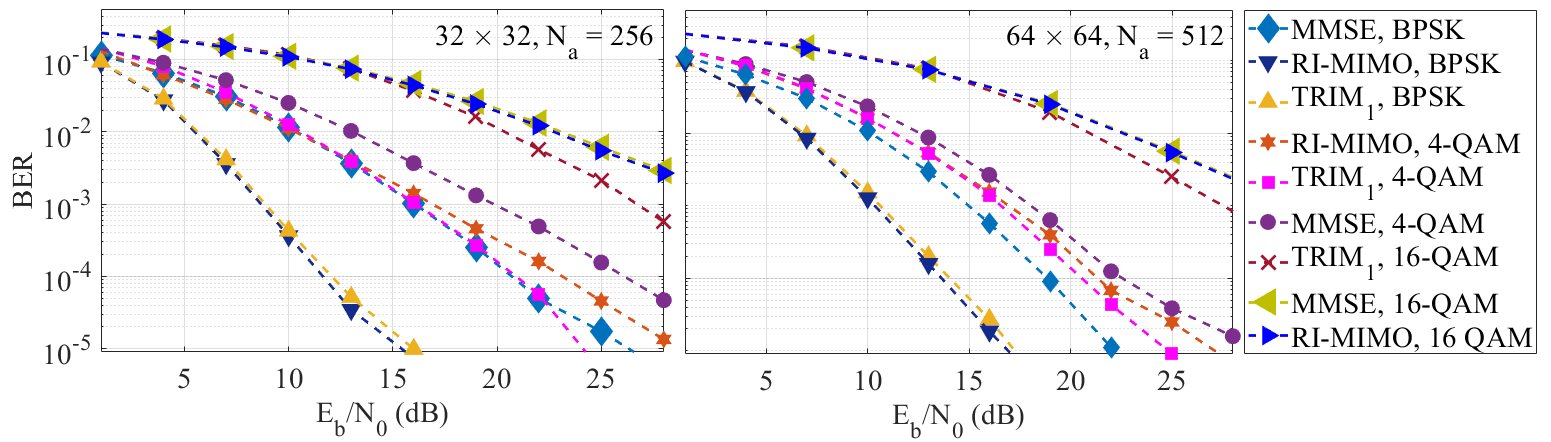}
 \caption{Bit Error Rate curves for $32\times32$ and $64\times64$ MIMO systems with BPSK, 4-QAM and 16-QAM modulations, illustrating the performance of RI-MIMO and TRIM on CIM.}
\label{fig:16x16_RI_V_LARGE}
\end{figure*}
In this section, we will discuss the tuning of the value for the regularisation prefactor for \textit{RI-MIMO} in Eq.~\ref{eq:ri-mimo}, $r(\rho,M,N_t)$, relative to the magnitude of Ising coefficients of the original un-regularised problem. To maintain consistency of results, we normalize the Ising coefficients of the original problem to $[-1,1]$. Starting from the $16\times16$ BPSK baseline MIMO system, we compute performance for various values. 

In order to determine the impact of modulation($M$), number of users ($N_t$) and SNR ($\rho$)
on the optimal value of $r(\rho,M,N_t)$, we look at BER vs regularisation factor for various MIMO sizes and modulations, while keeping SNR fixed at 10 dB and 15 dB in Fig.~\ref{fig:reg_v_numMod}. We note that the BER reduces dramatically from $r = 0$ (unregularized) to around $r = 0.1$, beyond which the sensitivity of BER to choice of $r$ is not much. We note that the optimal value is around 0.15, which acts as a threshold: for larger $r$ the BER performance is only slightly affected. Based on these observations, for practicality, in our benchmarks, we will be using $r(\rho,M,N_t) = 0.15$, irrespective of SNR, modulation, and the
number of users. In a practical system, similar experiments can be used to construct a lookup table for the optimal value of $r$ as a function of $M,N_t$ and $\rho$.

Using the prescriptions above, Fig.~\ref{fig:16x16_higherMod} provides the BER performance of \textit{RI-MIMO} and \textit{TRIM} for a $12\times12$ and $16\times16$ MIMO system with 4 QAM and 16 QAM modulations. 
We see that the error floor problem appears even for higher modulation, and \textit{RI-MIMO} successfully mitigates it. Note that the difference between \textit{RI-MIMO} and MMSE reduces as the modulation order increases. However, TRIM provides superior performance for the same number of total anneals and provides 10-15 dB gain over MMSE. Finally, we expand our evaluation to have even more antennas and users ($32\times32$ and $64\times64$ MIMO) in Fig.~\ref{fig:16x16_RI_V_LARGE} and see similar performance trends for TRIM, RI-MIMO and MMSE as observed for our baseline $16\times16$ MIMO.  
\subsection{Finite Precision of Ising Machines}
\label{sec:finitePrecision}
\begin{figure*}[h]
    \centering
    \includegraphics[width=\linewidth]{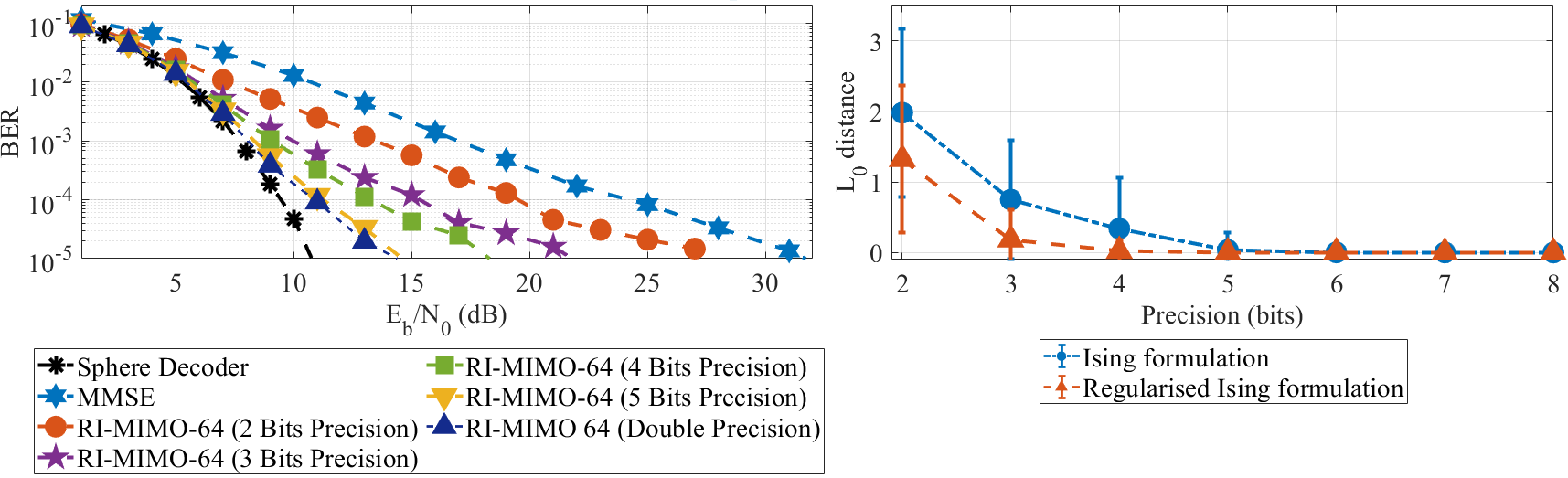}
    \caption{(left) Bit Error Rate Curves for 16$\times$16 MIMO and BPSK modulation, illustrating the performance of RI-MIMO on Coherent based Ising Machines (CIM) with varying precision. 
    (right) $L_0$ deviation from the ground state for $8\times8$ MIMO and BPSK modulation.
    }
    \label{fig:16x16_RI_Precision}
\end{figure*}
\begin{figure*}
    \centering
    \includegraphics[width=\linewidth]{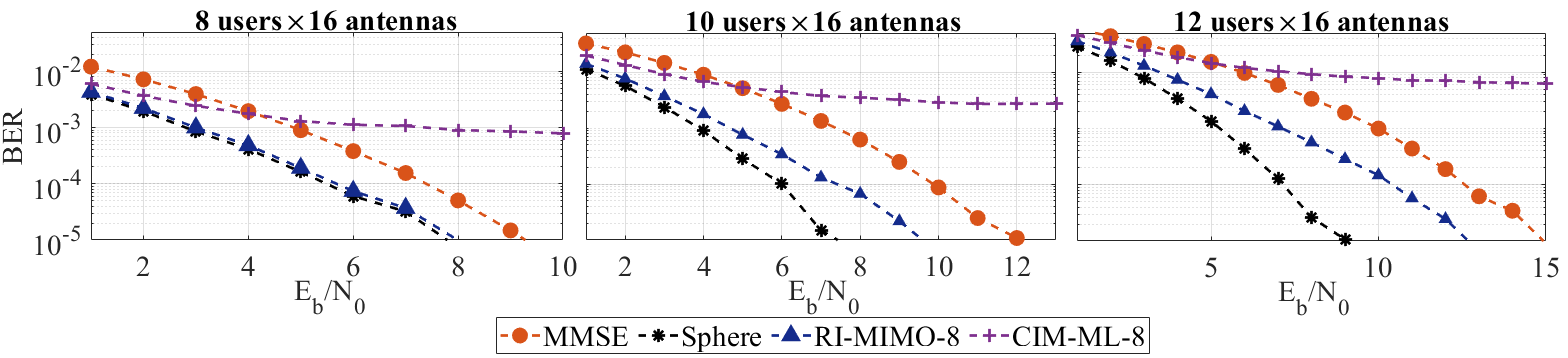}
    \caption{Bit Error Rate Curves for various Massive MIMO system with BPSK modulation and 16 antennas at the base station, illustrating the performance of RI-MIMO. 
    }
     \label{fig:16x16_massive}
\end{figure*}
\begin{figure*}
\centering
    \includegraphics[width=\linewidth]{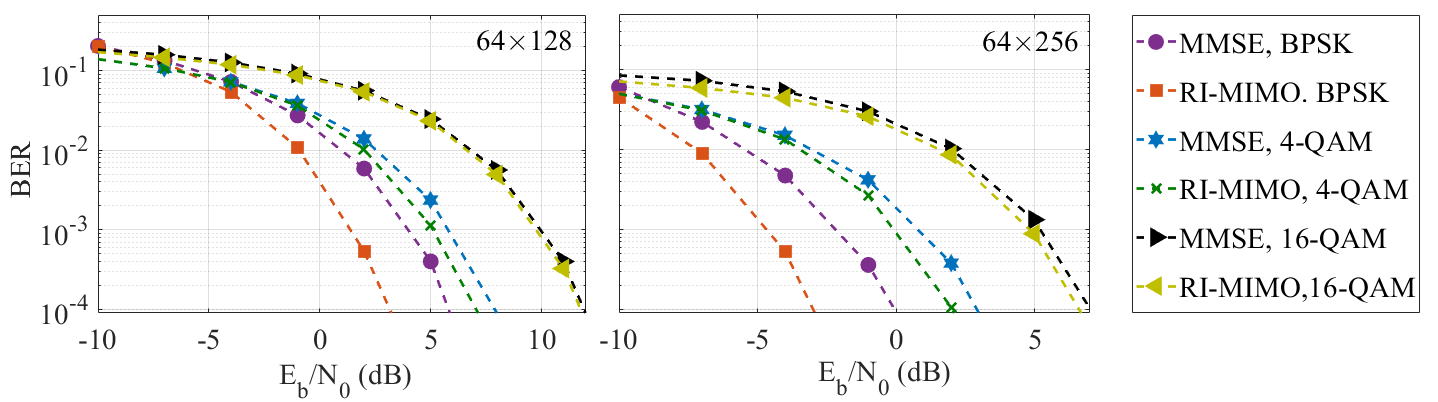}
    \caption{BER comparison of MMSE and RI-MIMO-512 for massive MIMO systems with very large number of antennas at the base station.}
    \label{fig:massiveMIMO_largeBER}
\end{figure*}
Physical realizations of CIMs typically involve analog control of the couplings between spins. Consequently, the precision in specifying the coefficients of an Ising problem ($J_{ij}$) can be limited. Currently, published results of optical hardware implementations of CIMs have only been for problems where the Ising coefficients take values $\{-1,0,1\}$, and even digital optimized implementations on GPUs are limited to floats encoded with $\approx5$ bits. While the precision may be improved, any analog machine will have a rather strict practical limit on how much precision can be achieved. Current quantum annealers have a similar limit on specifying Ising coefficients accurately, which can lead to drastic deterioration of performance~\cite{jchaos}. In this subsection, we look at the performance of \textit{RI-MIMO} under the constraint that Ising coefficients can be expressed using $K$ bits only. All Ising coefficients are normalized to $[-1,1]$; one bit is used to indicate the sign (positive/negative), and $K$-1 bits are used to express the decimal part of the Ising coefficient. 
We simulate \textit{RI-MIMO} with limited precision CIM for the baseline $16\times16$ BPSK MIMO system in Fig.~\ref{fig:16x16_RI_Precision}(left). We see that 5 bits of precision are enough for achieving performance similar to double precision for the given system. If we reduce precision below 5 bits, then the performance starts to deteriorate. It is interesting to note that, even with 2-bit precision, where the Ising coefficients can take values $\{-1,0,1\}$ only, \textit{RI-MIMO} performs much better than \textit{MMSE}. Note that for a more complicated system with more users or higher modulations, the precision requirement might be higher - but we leave the analysis to a future study. Independently from the optimization solver, we could study the "Resilience"~\cite{resil} of a problem as a metric to study the changes in ground state configuration due to noise in representing Ising coefficients. In Fig.~\ref{fig:16x16_RI_Precision}(right), we look at the $L_0$ distance between the original ground state and the ground state of the Ising problem represented with finite precision. It is a natural metric for wireless applications; as for BPSK, it essentially represents the BER. In Fig.~\ref{fig:16x16_RI_Precision}(right), we observe for an indicative $8\times8$ BPSK instance set that $L_0$ deviation is nearly zero for 5 bits of precision and that the regularized Ising formulation is more resilient than the unregularized.
\begin{figure*}
    \centering
    \includegraphics[width=\linewidth]{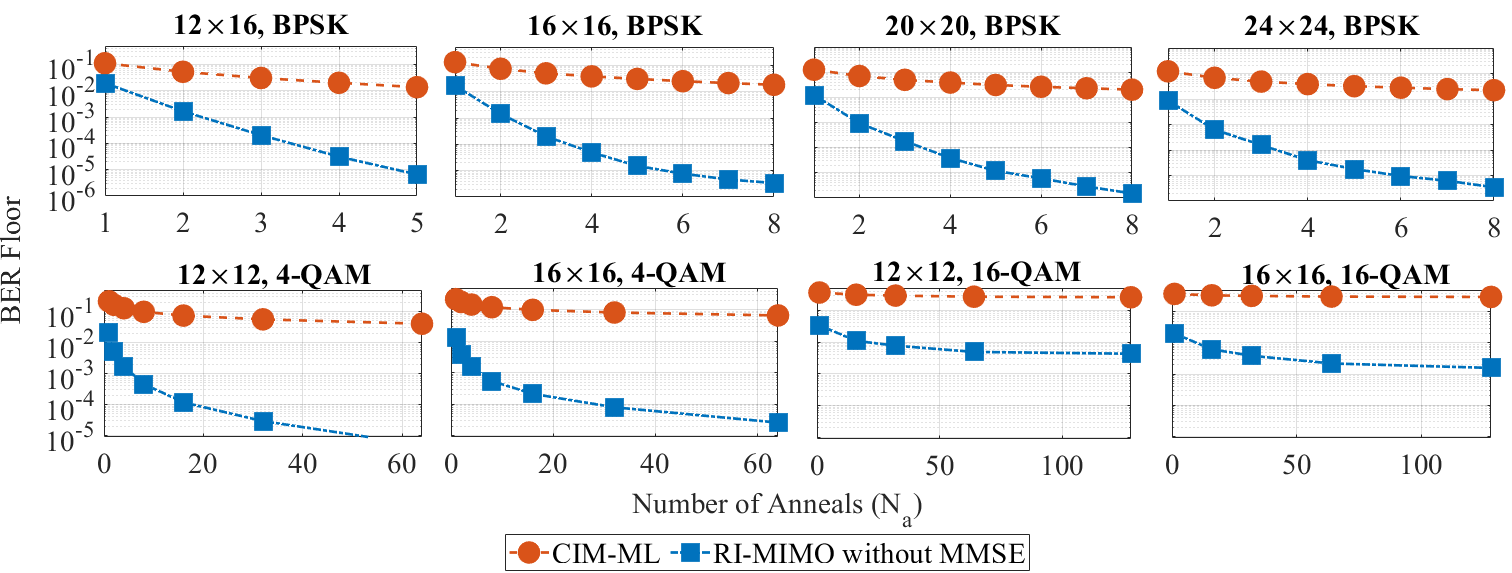}
    \caption{Variation of Error Floor with the number of anneals if RI-MIMO is run without considering MMSE solution as a candidate for various MIMO systems, illustrating that it takes only a few anneals to reduce the resultant error floor to the very low values. In contrast, the error floor associated with CIM-ML reduces very slowly with an increasing number of anneals.}
    \label{fig:rimimo_eff}
\end{figure*}
\subsection{Massive MIMO}
\label{sec:MassiveMimo}
In this section, we will address the need and performance of RI-MIMO/TRIM for massive MIMO systems. Massive MIMO systems tend to have a much higher number of antennas at the receiver than at the transmitter. Due to this, the channel is extremely well-conditioned, and even linear detectors such as MMSE or iterative algorithms can perform extremely well~\cite{massiveMimoNearOpt}. However, having a large number of antennas at the base station and yet supporting a small number of users reduces the overall cell throughput. Hence, the low BER of linear methods comes at the price of the total number of users served. We will see that for massive MIMO systems, where the linear methods are not optimal, RI-MIMO can be used to obtain optimal performance with very low complexity. In Massive MIMO systems, where the ratio between the number of receiver antennas and the number of transmitter antennas is three or more, MMSE can provide performance similar to the Sphere Decoder. However, the total cell throughput achieved is much lower than what may be possible, with near-optimal MIMO detection, in an equivalent large MIMO system (where the number of receiver antennas and transmitter antennas is the same).  In Fig.~\ref{fig:16x16_massive} we look at massive MIMO systems with 16 antennas at the base station and BPSK modulation. We see that for $8\times16$ and $10\times16$ MIMO with BPSK, the MMSE decoder is sub-optimal, and RI-MIMO can provide near-optimal performance at very low complexity. In Fig.~\ref{fig:massiveMIMO_largeBER}, we expand our evaluation to more realistic massive MIMO scenarios with 128 and 256 antennas at the base station using BPSK, 4-QAM, and 16-QAM modulations. Note that, for such systems, we are unable to compute the optimal BER due to extremely large complexity of the Sphere decoder. We note that, RI-MIMO performs much better than MMSE for such systems when BPSK or 4-QAM modulation is used. However, as noted before, the performance gains are much less for 16-QAM modulation. Improving the performance of our methods for 16-QAM and higher modulations is the key focus of our future work.

Before we go ahead and look at spectral efficiency, note that if we do not consider the MMSE solution as a candidate, then RI-MIMO will have an error floor as well. However, due to a much higher probability of success than the conventional approach, the error floor associated with RI-MIMO will be much lower and reduced much more quickly, to very low values, in comparison to the conventional approach. We see from Fig.~\ref{fig:rimimo_eff} that it take just 4-6 anneals per instance (much less than the conventional approach) for a $16\times16$, $20\times20$, and $24\times24$ MIMO system with BPSK modulation, to reduce the error floor to $10^{-5}$. In contrast, the error floor associated with CIM-ML reduces very slowly, and even with 64 anneals per instance, it remains around $10^{-3}$ (as seen in Fig~\ref{fig:16x16_RI_OIM_QA}). If we consider the MMSE solution as a candidate, then there is no error floor, as MMSE will make sure that BER goes to zero as SNR goes to infinity. 
\begin{figure*}[h!]
\centering
    \includegraphics[width=\linewidth]{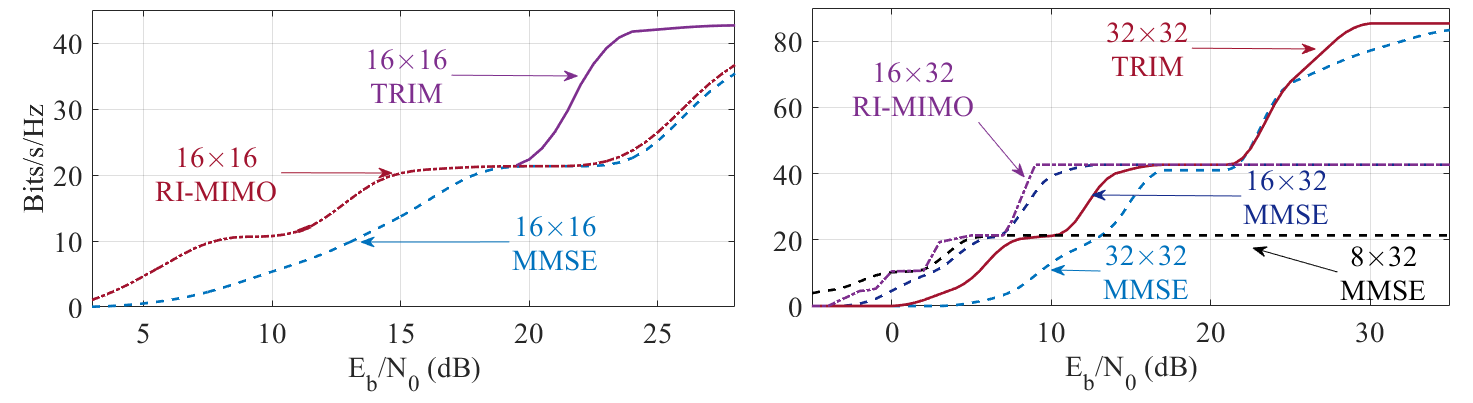}
    \caption{Spectral Efficiency: Comparison between throughput of MMSE, TRIM, RI-MIMO (using adaptive MCS for all) in large/massive MIMO scenarios: (left) 16 antennas at the base station and we use $N_a = 64$, (right) 32 antennas at the base station and we use $N_a = 256$.}
    \label{fig:16x16_amc_trim}
\end{figure*}
\begin{figure*}
\centering
    \includegraphics[width=\linewidth]{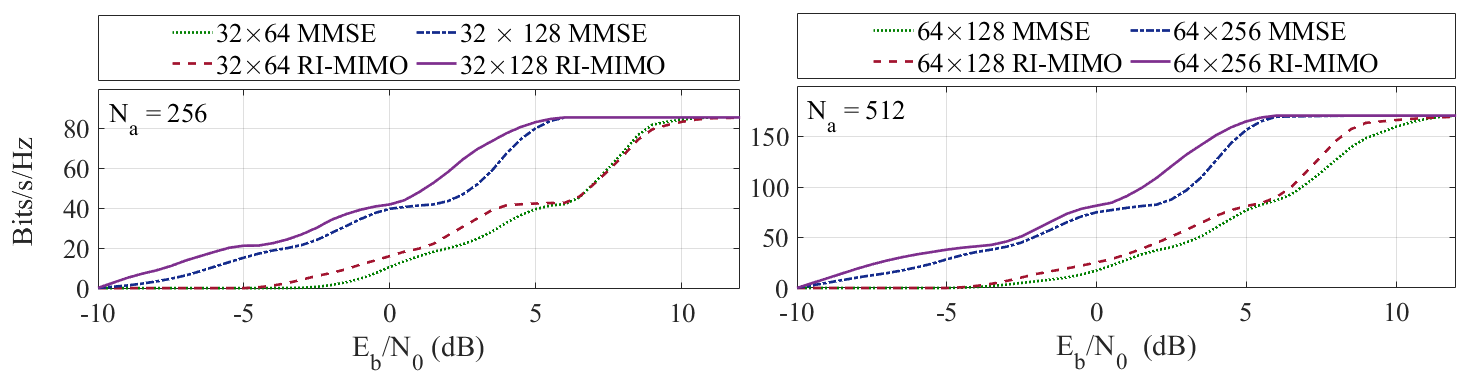}
    \caption{Spectral Efficiency: Comparison between throughput of MMSE and RI-MIMO for massive MIMO scenarios with large number of antennas at the base station. }
    \label{fig:massiveMIMO_tput}
\end{figure*}
\subsection{Throughput with MCS adaptation}
In this section, we will try to look at the spectral efficiency of RI-MIMO. In a practical system, the Modulation and Coding Scheme (MCS) associated with the data transmission is selected dynamically based on the SNR experienced by the user. Usually, the BS has a list fixed set of (modulation, coding rate) and selects the suitable tuple based on Channel State Information (CSI) measurements. To emulate such a system, we consider several MIMO systems that use feed-forward convolutional coding with 1500-byte packets for data transmission. 

Usually, there is an Adaptive Modulation and Coding (AMC) module that selects a suitable MCS based on the CSI measurements to maximize the throughput performance. For this section, we ignore the complexities of the AMC module and assume that there is an oracle AMC that always selects the best MCS such that the overall throughput is maximized at any given SNR. The AMC module can select among BPSK, 4-QAM, or 16-QAM modulation, and among $\dfrac{1}{3}$, $\dfrac{1}{2}$ or $\dfrac{2}{3}$ convolutional coding rate. 
 \begin{figure*}[h!]
 \centering
    \includegraphics[width=\linewidth]{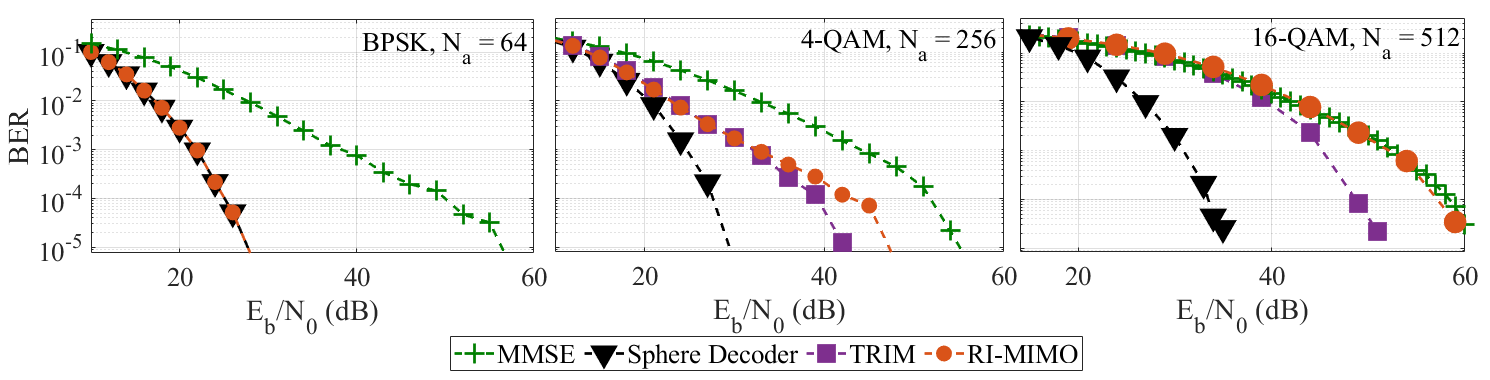}
    \caption{Bit Error Rate for $8\times8$ MIMO with QuaDRiGa channels, demonstrating that the performance trends for various tested algorithms are similar to the Rayleigh fading experiments.}
    \label{fig:ber_quadriga}
\end{figure*}
\begin{figure*}[h!]
    \centering
    \includegraphics[width=\linewidth]{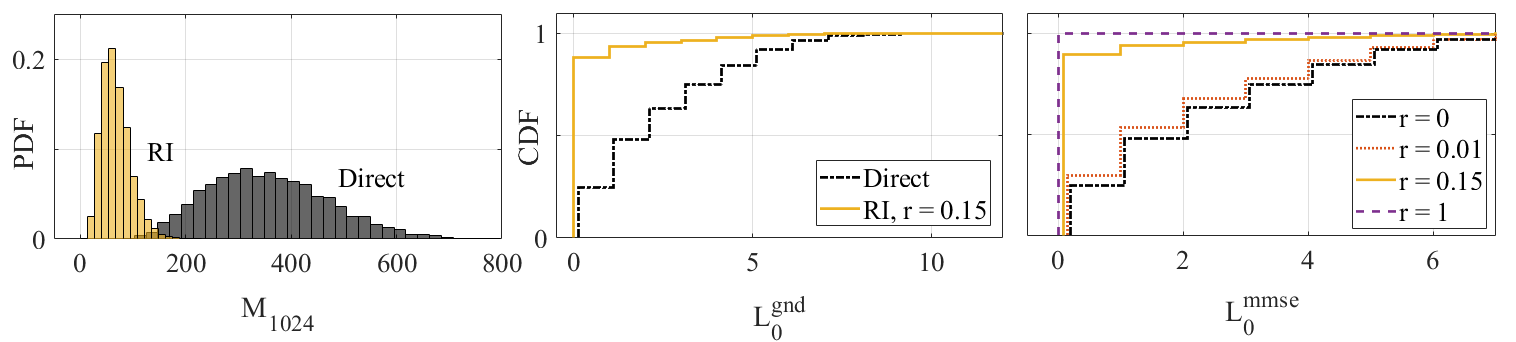}
    \caption{Empirical distribution of the number of locally stable states and their $L_0$ distance from the ground state and the MMSE solution for $16\times16$ MIMO, BPSK, and 10 dB SNR.}
    \label{fig:minimaSuccess}
\end{figure*}
We see from Fig~\ref{fig:16x16_amc_trim}(left) that \textit{RI-MIMO} and \textit{TRIM} allows us to operate using aggressive modulations and coding schemes and hence achieve much better performance. In particular \textit{RI-MIMO} achieves around 2.5$\times$ more throughput in low-SNR regime ($\approx 7.5 dB$) and 2$\times$ more throughput in mid-SNR regime( $>$ 15dB). In the high-SNR ( $>$ 20dB) both MMSE and \textit{RI-MIMO}-64 seem to provide similar throughput, because \textit{RI-MIMO}-64 is not sufficient for 16 QAM modulation (as noted before). However, TRIM allows us to get good performance with 16 QAM and provides a 2$\times$ throughput gain in the high-SNR regime. With Increasing SNR, the channel capacity also increases; hence we would expect the AMC module to use more aggressive modulations in order to achieve the best possible capacity.
Consequently, improving the performance of RI-MIMO/TRIM for 16 QAM and higher modulations remains a key challenge for future work. Further, we simulate several massive MIMO systems in Fig~\ref{fig:16x16_amc_trim}(right). Although massive MIMO systems with a high ratio between the number of antennas at the base station and the number of users, like $4\times16$, can achieve near-optimal performance with just the MMSE receiver, this simplicity comes at the cost of limiting the overall cell throughput. We see from Fig~\ref{fig:16x16_amc_trim}(right) that with RI-MIMO/TRIM, BS can expand these systems to support more users concurrently and drastically increase the overall cell throughput. We see from Fig~\ref{fig:16x16_amc_trim}(right) that RI-MIMO/TRIM allows us to expand a $8\times32$ to $16\times32$ or $16\times32$ to $32\times32$ MIMO and increase the overall cell throughput by twice. Note that, even though the spectral efficiency of MMSE and TRIM is similar in high-SNR conditions for $32\times32$ MIMO, it is not feasible to expand $16\times32$ MIMO to $32\times32$ MIMO with just MMSE because, unlike TRIM, the performance of MMSE is terrible in low SNR situations. In Fig.~\ref{fig:massiveMIMO_tput}, we simulate massive MIMO scenarios which resemble real-world deployments with 64, 128, and 256 antennas at the BS. We see that, RI-MIMO can provide significant throughput gains over MMSE in the low-SNR and mid-SNR regimes. We also observe that throughput of RI-MIMO is similar to MMSE in the high-SNR conditions, due to the previously noted limitation of RI-MIMO in performing well with 16-QAM.
\subsection{Evaluation on QuaDRiGa Channel Model}
In this section, we evaluate our methods on more realistic channels generated by the QuaDRiGa simulator~\cite{quadriga}. The network topology is as follows: a single base station with an "ula8" antenna array consisting of eight antennas at the center of a hexagonal cell situated at the height of 10m. The channel instances are calculated assuming the users are distributed randomly in a radius of 50m around the base station. Each user is equipped with a single "Omni" antenna at a height of 1.5m. The system operates at a frequency of 2.6GHz, and we assume flat fading channels. We observe in Fig.~\ref{fig:ber_quadriga}, that RI-MIMO/TRIM provides significant gains over MMSE over the QuaDRiGa channels as well. We further note that the performance gains of RI-MIMO are severely affected for 16-QAM, as noted previously as well, and improving the performance for 16-QAM and higher modulations remains the key focus of our future work. These observations are consistent with our previous experiments using Rayleigh fading channels. Although the exact BER characteristics might differ due to changes in the statistical properties of the channel, we do not observe any significant changes in the relative performance of different methods.
\subsection{Statistical Comparison of Direct and Regularised Ising Formulations}
In this section, we will empirically study the distribution of solutions returned by our method. In Fig.~\ref{fig:minimaSuccess}, we simulate $16\times16$ MIMO, BPSK modulation, 10 dB SNR, and $N_a = 1024$. As all stochastic optimization algorithms, we are sensitive to the existence and number of locally optimal states in the cost function landscape. For a given MIMO instance, characterized by $(\mathbf{H},\mathbf{y})$, we define $M_k = $ the number of distinct states returned by the CIM machines over $k$ anneals. Note that the ground state may or may not be a part of these $M_k$ states. $M_k$ is a random variable over the MIMO instances, and we look at the distribution of $M_{1024}$ in Fig~\ref{fig:minimaSuccess}(left); we see that our proposed RI formulation drastically reduces the number of locally stable states returned by the CIM solver. Intuitively this could be at the origin of the observed decrease in the probability of getting stuck in a local minimum. Next, we look at the ``quality" of various local minima in terms of how many of its spins differ from the ground state, \emph{i.e.}, the $L_0$ Hamming distance from the ground state~($L_0^{gnd}$). In Regularised Ising~(RI) formulation, we penalize the deviations from the MMSE solution with the goal of increasing the energy of locally stable states that are ``far away". This is indeed what we observe in Fig.~\ref{fig:minimaSuccess}(center); we see that locally stable states of the RI formulation are more likely to be closer to the ground state, and the probability of success (equal to the probability of finding a solution at zero distance) is much higher for the RI formulation. As a result, both the frequency of finding the ground state and the quality of locally stable states is higher for RI formulation, which leads to a better BER performance and a much lower BER floor (as seen in Fig.~\ref{fig:rimimo_eff}). In Fig~\ref{fig:minimaSuccess}(right), we look at the impact of tuning the regularization factor $r(\rho,M,N_t)$. The regularisation term penalizes the deviations from the MMSE solution, and a high regularisation factor should limit the search around the MMSE solution only. We observe the same in Fig~\ref{fig:minimaSuccess}(right); as $r$ increases, the solutions obtained by the CIM are closer to the MMSE solution. However, choosing $r$ to be very high will limit the search to the MMSE solution, and CIM will miss the ground state more often; our empirically optimized choice of $r = 0.15$ strikes a balance and optimizes for the best BER. 

%% file: conclusion.tex
\section{Conclusion}
\label{sec:conclusions}
In this paper, we explore the application of Coherent Ising Machines (CIM) for maximum likelihood detection for MIMO detection. We see that previous approaches used by MIMO detectors based on the Ising model suffer from an error floor problem and, unless many repetitions are allowed, does not have a satisfactory Bit Error Rate (BER) performance in practice. We propose a novel Regularized Ising approach and show that it mitigates the error floor problem and is a viable method to be implemented on the hardware implementation of Coherent Ising Machines. We propose two MIMO detection algorithms based on regularised Ising approach (\textit{RI-MIMO} and \textit{TRIM}). We demonstrate, using a CIM simulator, that our algorithms can outperform the previous Ising approach and have the potential to achieve near-optimal performance for large MIMO systems. 

While our experiments use a CIM simulator, we have identified the constraints that need to be satisfied by a physical implementation of the CIM to meet the processing requirements of an LTE system. As a next step, we plan to evaluate our methods on more recent extensions to the CIM (e.g., amplitude-heterogeneity correction~\cite{leleu2019destabilization}) and on an experimental CIM implementation~\cite{mcmahon2016fully}.

In conclusion, our results are complementing the mounting evidence from generic spin-glass benchmarks~\cite{sankar2021benchmark, hamerly2019experimental, 100kCIM}, further indicating that Coherent Ising Machines (using the proposed Regularized Ising approach) are a promising candidate for providing a superior alternative to the existing MIMO detection approach and achieving near-optimal performance for practical systems with a large number of users and antennas, although performance and engineering aspects need to be substantially improved for practical deployment, especially for higher modulations.

%% file: acknowledgements.tex
\section*{Acknowledgments}

D.V. acknowledges support from NSF award CNS-1824470 and both D.V. and P.L.M. acknowledge support from NSF award CCF-1918549.
A.K.S. and K.J. acknowledge support from NSF award CNS-1824357. A.K.S.'s internship has also been supported by the USRA Feynman Quantum Academy, as part of the NASA Academic Mission Services (NNA16BD14C) -- funded under SAA2-403506. P.L.M. also acknowledges financial and technical support from NTT Research and membership in the CIFAR Quantum Information Science Program as an Azrieli Global Scholar.